\documentclass[english]{paper}
\usepackage[T1]{fontenc}
\usepackage[latin9]{inputenc}
\usepackage{float}
\usepackage{units}
\usepackage{amsmath}
\usepackage{amsthm}
\usepackage{amssymb}
\usepackage{graphicx}
\usepackage{csquotes}
\usepackage[backend=biber, style=nature]{biblatex}
\newcommand{\noun}[1]{\textsc{#1}}
\numberwithin{equation}{section}
\numberwithin{figure}{section}

\usepackage{tikz}
\usepackage{tikz-3dplot}
\usetikzlibrary{arrows,external}
\usetikzlibrary{shapes,backgrounds,fit}

\usepackage{pgfplots}
\pgfplotsset{compat=1.3}
\usepackage[detect-family]{siunitx}

\usepackage[eulergreek]{sansmath}
\usepackage{relsize}

\usepackage{babel}

\begin{document}

\title{The acceptance of the HiSPARC experiment}

\author{N.G. Schultheiss\thanks{Nikhef} \thanks{Zaanlands Lyceum}} 
\maketitle
\begin{abstract}
Cosmic ray primary particles initiate extended air showers (EAS) in the atmosphere. 
The directions of these cosmic rays approximate a homogeneous isotropic distribution. 
The HiSPARC experiment, consisting of a growing number of measurement stations 
scattered over the Netherlands, Denmark and the United Kingdom, detects EAS particles 
using scintillators. These detections facilitate reconstructions of the direction of cosmic
ray primaries. 

The acceptance of the HiSPARC experiment, limited due to the generating mechanism of 
EASs in the atmosphere and the location of the measurement stations, has been analysed.
\end{abstract}

\begin{keywords}
Cosmic rays, acceptance, HiSPARC.
\end{keywords}

\section{Introduction}

A high energy primary cosmic ray particle (nucleus, proton, electron, gamma) arrives from a 
point defined by the celestial bound right ascension/declination $\left(\alpha,\delta\right)$-frame. 
The primary particle generates an EAS (extended air shower) in the atmosphere. HiSPARC 
stations detect particles in this EAS using two or four scintillators. The location of this station is 
defined by its longitude and latitude in an Earth bound $\left(\lambda,\phi\right)$-frame. 

The acceptance of each scintillator is expressed as a distribution of the number of detected EAS 
per direction  \cite[p. 78]{G-PCRP}, \cite{NGS-SP:arxiv-2016}. This direction is described using 
the local zenith/azimuth $\left(\theta,\varphi\right)$-frame. An east/west effect for high energy 
charged particles due to the Earth's magnetic field is negligible. So the distribution is assumed 
to be independent of the azimuth $\varphi$. 

Because the Earth rotates with respect to the celestial sphere, the transformation of the 
$\left(\theta,\varphi\right)$-frame via the $\left(\lambda,\phi\right)$-frame into the 
$\left(\alpha,\delta\right)$-frame is time dependant.

\section{The distribution of HiSPARC-stations}

HiSPARC measurement stations are maintained by participating schools and universities scattered 
over the Netherlands, Denmark and the United Kingdom, approximately at sea level. As a result 
longitudes and latitudes are not evenly distributed. The moment in time and the longitude define 
the mean right ascension of the measured EASs. The latitude defines the mean declination of 
measured EASs. 

A point in the sky above a hypothetical station located on the Equator is visible during a
maximum of 12 sidereal hours in a sidereal day because of the rotation of the Earth. 
The detection time decreases considerably due to the extinction of EASs for larger 
zenith angles.

A hypothetical station located on the North Pole also detects EASs arriving from the sky
above the horizon. Here detections are possible during the entire sidereal day. This 
increases the detection time dramatically. As a result higher latitudes lead to a 
longer time window for detections of EASs, and there corresponding cosmic rays, arriving 
from a smaller part of the sky.

\begin{figure}[H]
\noindent \begin{centering}
\includegraphics[width=8cm]{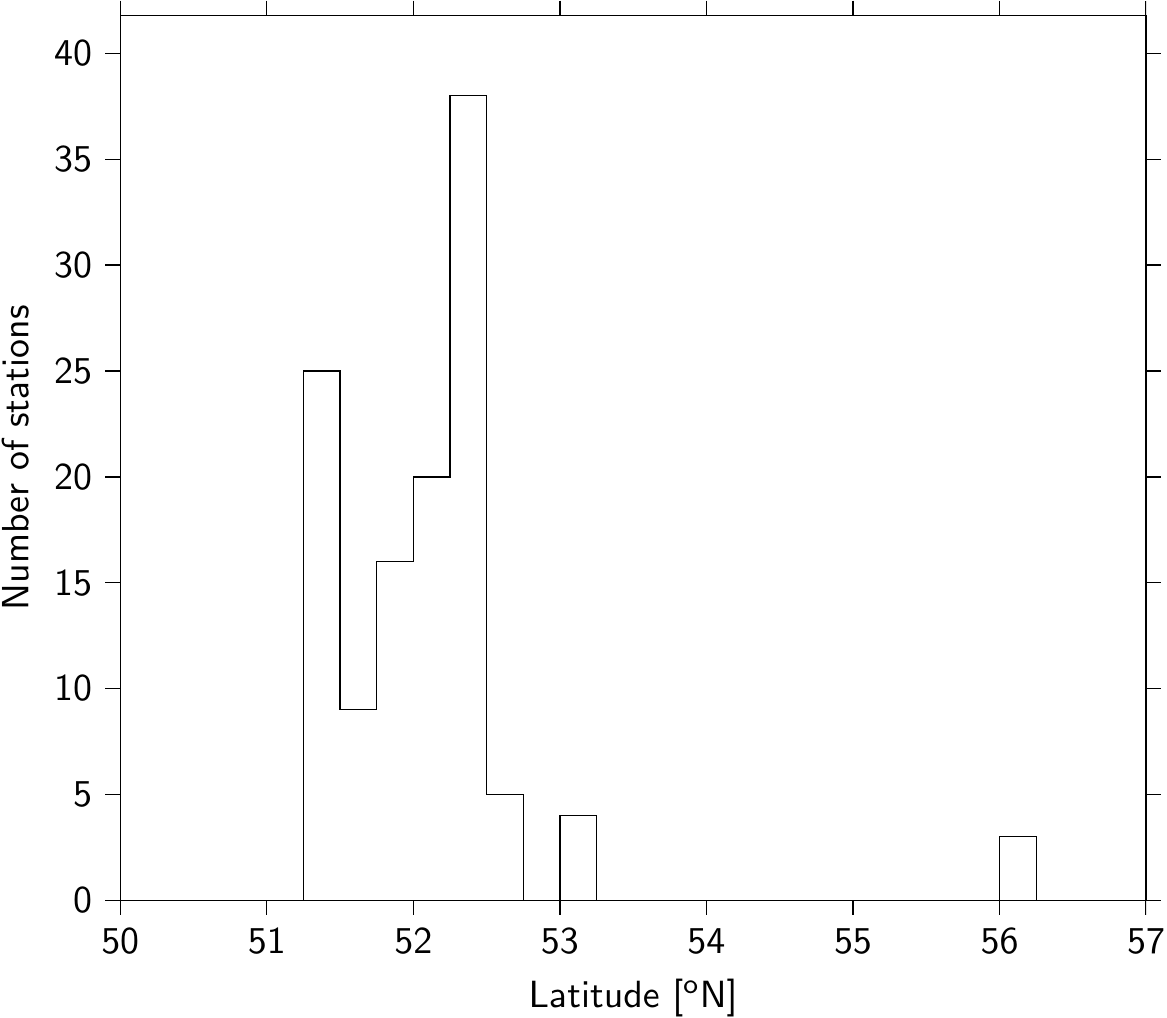}
\par\end{centering}

\caption{\label{fig:Latitudinal-distribution}\textit{The number of stations per latitude. The latitude defines
the width of the time window for EAS detections. The width increases
for higher latitudes and decreases for lower latitudes.}}
\end{figure}

\newpage

\begin{figure}[H]
\noindent \begin{centering}
\includegraphics[width=8cm]{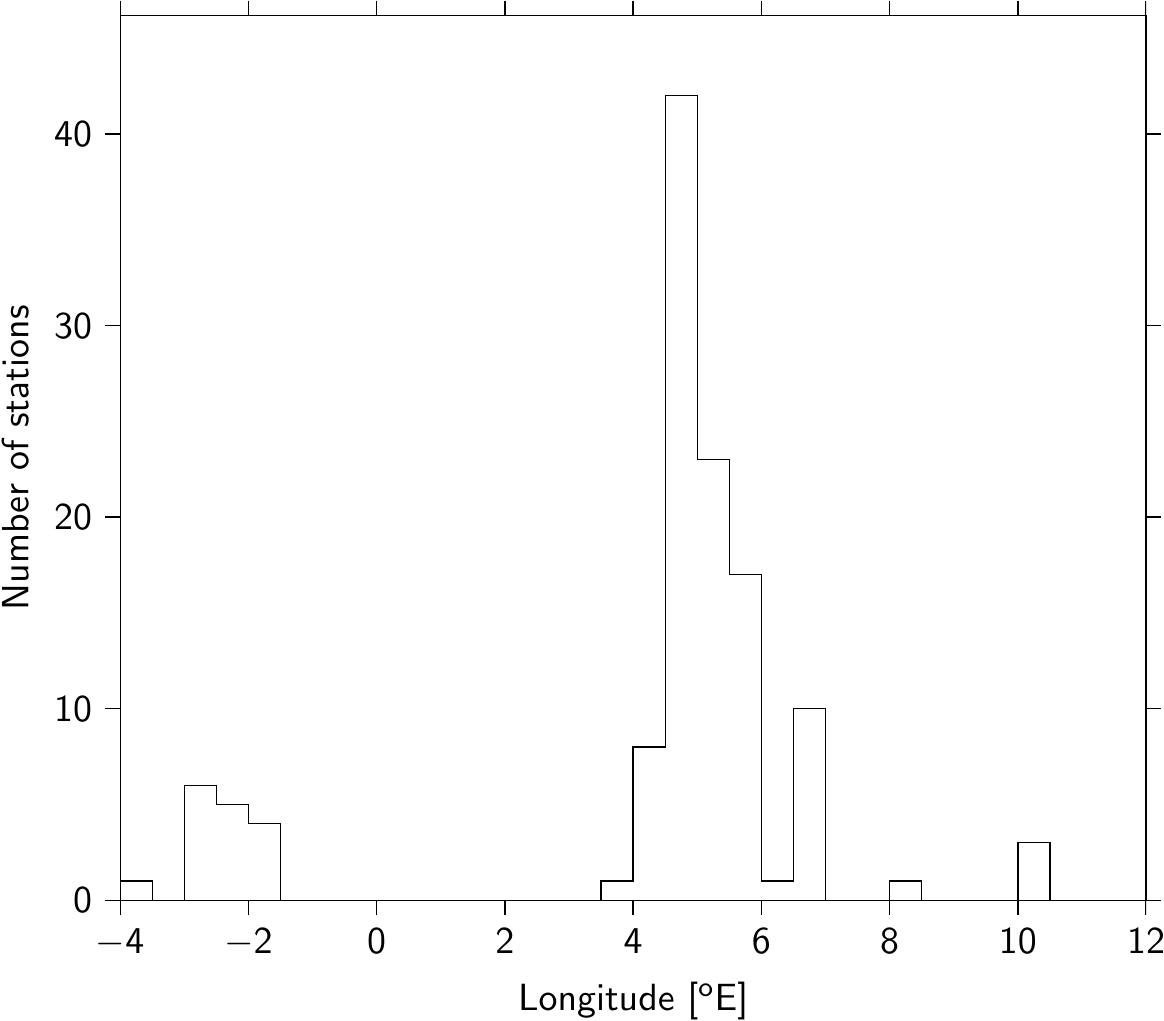}
\par\end{centering}

\caption{\label{fig:Longitudinal-distribution}\textit{The number of stations per longitude. Easterly stations 
detect cosmic rays coming from a celestial area before westerly stations.}}
\end{figure}

The longitudinal distribution of HiSPARC stations is shown in fig. \ref{fig:Longitudinal-distribution}. 
This distribution defines the center of the time window for detectable EASs. Longitudes are between 
$\ang{-3}$ and $\ang{10.5}$E. The Earth turns $\ang{15}$/hour. A station at $\ang{10.5}$E detects cosmic 
rays arriving from a celestial area about 54 minutes before a station at $\ang{-3}$E\footnote{A cosmic ray 
source is in transition, moving from the local eastern celestial hemisphere to the western hemisphere.}. 

\section{Zenith angle}

\begin{figure}
\noindent \begin{centering}
\includegraphics[width=4cm]{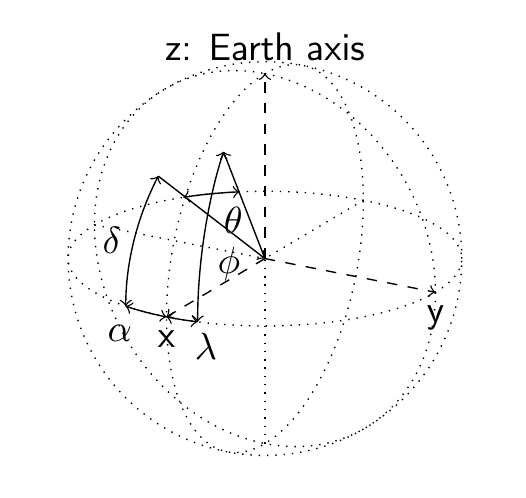}
\par\end{centering}

\caption{\label{fig:A-cosmic-ray}\textit{A cosmic ray primary hitting the
Earth, placed in the centre of the celestial sphere. Shown is the equatorial
frame. The zenith angle $\theta$ is the angle between a cosmic ray
primary from $\left(\alpha,\delta\right)$ in the (Right Ascension,
declination)-frame and the zenith of the detection station at $\left(\lambda,\phi\right)$
in the (longitude, latitude)-frame. The rotation of the Earth results
in a time dependent transformation from the $\left(\lambda,\phi\right)$-frame
to the $\left(\alpha,\delta\right)$-frame.}}
\end{figure}

Fig. \ref{fig:A-cosmic-ray} shows the zenith angle between the traveling direction of the cosmic ray primary 
and the zenith of the detection station located at $\left(\lambda,\phi\right)$. The location of the detection 
station on a given time is expressed in the ECEF\footnote{Earth Centre Earth Fixed, the equatorial 
plane is defined by $x$and $y$. The $x$-axis points towards the Greenwich meridian. TheEarth axis, 
pointing to Polaris, defines the $z$-axis.}-frame using the longitude $\lambda$ and latitude $\phi$ as: 

\begin{equation}
\left(\begin{array}{c}
x_{s}\\
y_{s}\\
z_{s}
\end{array}\right)=\left(\begin{array}{c}
\cos(\phi)\cos(\lambda)\\
\cos(\phi)\sin(\lambda)\\
\sin(\phi)
\end{array}\right)
\end{equation}

\newpage

The direction of the cosmic ray induced air shower can be expressed on a given time in the ECEF-frame 
using the right ascension $\alpha$ and the declination $\delta$. 

\begin{equation}
\left(\begin{array}{c}
x_{c}\\
y_{c}\\
z_{c}
\end{array}\right)=\left(\begin{array}{c}
\cos(\delta)\cos(\alpha)\\
\cos(\delta)\sin(\alpha)\\
\sin(\delta)
\end{array}\right)
\end{equation}

Using these expressions the zenith angle $\theta$ is the angle between the direction of 
the cosmic ray and the location of the scintillator:

\begin{equation}
\cos\left(\theta\right)=\left(\begin{array}{c}
\cos(\delta)\cos(\alpha)\\
\cos(\delta)\sin(\alpha)\\
\sin(\delta)
\end{array}\right).\left(\begin{array}{c}
\cos(\phi)\cos(\lambda)\\
\cos(\phi)\sin(\lambda)\\
\sin(\phi)
\end{array}\right)\label{eq:innerProduct}
\end{equation}

A detection station, located on the Earths surface, turns $360^{\mathrm{o}}$ around the 
$z$-axis in a sidereal day. The Right Ascension is the only time dependent variable. 
Using the local hour angle\footnote{The local hour angle of an area is defined as the angle 
between the area and the zenith, both projected on the equatorial plane. One hour is equal 
to an angle of $15^{\mathrm{o}}$.} $\alpha_{l}$ instead of the right ascension, the local 
longitude of the scintillator is set to $\lambda=0^{\mathrm{o}}$:

\begin{equation}
\cos\left(\theta\right)=\left(\begin{array}{c}
\cos(\delta)\cos(\alpha_{l})\\
\cos(\delta)\sin(\alpha_{l})\\
\sin(\delta)
\end{array}\right).\left(\begin{array}{c}
\cos(\phi)\\
0\\
\sin(\phi)
\end{array}\right)
\end{equation}

\newpage

Leading to:

\begin{equation}
\cos\left(\theta\right)=\cos(\delta)\cos(\phi)\cos(\alpha_{l})+\sin(\delta)\sin\left(\phi\right)\label{eq:cosZenLatDec}
\end{equation}

The flux\footnote{It must be noted that the geometry of the detector leads to a dependency
of the distribution of $\cos\left(\theta\right)$. The extinction
of the shower on sea level is thus defined by $\cos^{\left(a-1\right)}\left(\theta\right)$.
The exponent $\left(a-1\right)$, defining the extinction, is energy
and height dependent.} in an area at $\left(\theta,\varphi\right)$ with a width of $dw=\sin\left(\theta\right)d\varphi$
and a height $d\theta$ is defined as \cite{NGS-SP:arxiv-2016}:

\begin{equation}
\frac{d^{2}N\left(\theta,\varphi\right)}{d\theta dw}\varpropto\cos^{a}\left(\theta\right)\label{eq:distribution_zenith}
\end{equation}

Substitution of eq. \ref{eq:cosZenLatDec} into eq. \ref{eq:distribution_zenith} leads to:

\begin{equation}
\frac{d^{2}N\left(\theta,\varphi\right)}{d\theta dw}\varpropto\left(\cos(\delta)\cos(\phi)\cos(\alpha_{l})+\sin(\delta)\sin\left(\phi\right)\right)^{a}\label{eq:rossi-local}
\end{equation} 

The local right ascension $\alpha_{l}$ is the time dependent variable.

\section{The longitudinal dependency of a detected celestial area}

Using eq. \ref{eq:rossi-local} the flux of cosmic rays is calculated as a function of the local hour angle and the 
location. The time window for an celestial area passing through the zenith of the detection station is calculated. 

\begin{figure}[H]
\noindent \begin{centering}
\includegraphics[width=4cm]{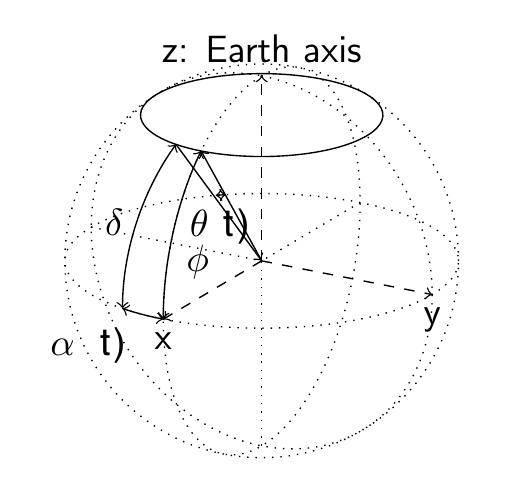}
\par\end{centering}

\caption{\label{fig:Delta-distribution}\textit{The longitudinal dependency
follows the distribution in a circle of constant declination trough
the zenith ($\delta=\phi$).} }
\end{figure}

In this case latitude and declination are equal, $\delta=\phi$. Eq.
\ref{eq:rossi-local} is written as: 

\begin{equation}
\frac{d^{2}N\left(\theta,w\right)}{d\theta dw}\varpropto\left(\cos^{2}(\delta)\cos(\alpha_{l})+\sin^{2}(\delta)\right)^{a}
\end{equation}

\begin{figure}[H]
\noindent \begin{centering}
\includegraphics[width=8cm]{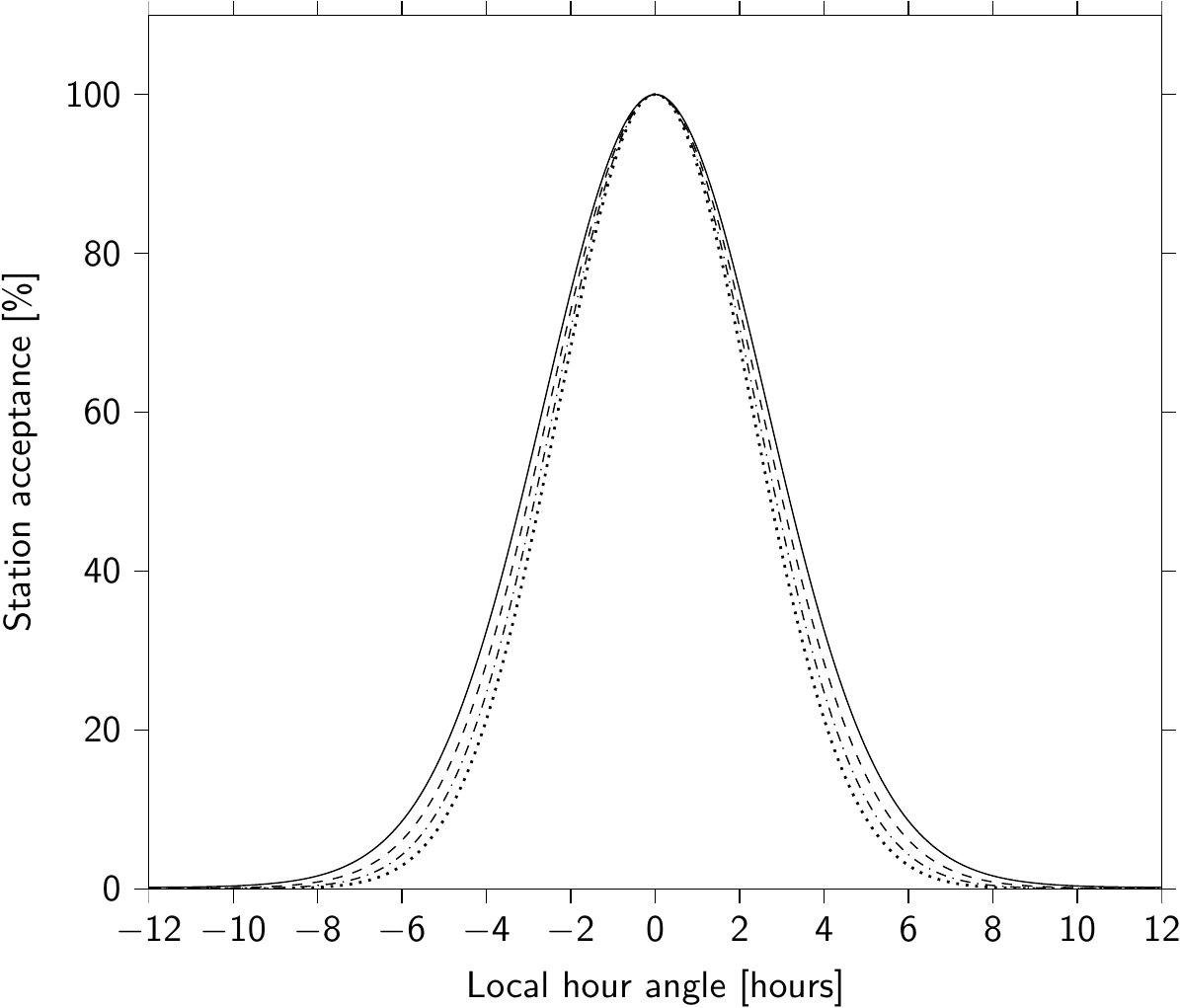}
\par\end{centering}

\caption{\noun{\label{fig:The-longitudinal-distribution}}\textit{The measured
time window depends on the declination. The distributions for $\delta=\phi$
are shown for the latitudes of $\phi=51^{\mathrm{o}}$ (dotted), $53^{\mathrm{o}}$
(dot dashed), $55^{\mathrm{o}}$ (dashed), and $57^{\mathrm{o}}$
(solid).}\textit{\noun{ }}\textit{On a local Right Ascension of 0
hours, when the celestial point is in transition and in the zenith,
the maximum acceptance of the station is reached ($100\%$). }}
\end{figure}

The location of a celestial area is defined in the $\left(\alpha_{l}\left(t\right),\delta\right)$-frame,
and moves along a circle of latitude ($\phi=\delta$) as show in fig.
\ref{fig:Delta-distribution}. The calculated EAS-flux from this area
is plotted in fig. \ref{fig:The-longitudinal-distribution} as a function
of the hour angle. EAS coming from this direction can roughly be measured
between 5 hours before and after the transition. 

The measured time window per station increases slightly when comparing
a northern with a southern stations within HiSPARC\footnote{The circumference of 
circles where latitude and declination are equal, decreases when the declination increases.}.
As shown in fig. \ref{fig:Latitudinal-distribution} the difference
in latitudes is much smaller than the field of view of a scintillator.

\section{Amsterdam Science Park}

\begin{figure}
\noindent \begin{centering}
\includegraphics[width=4cm]{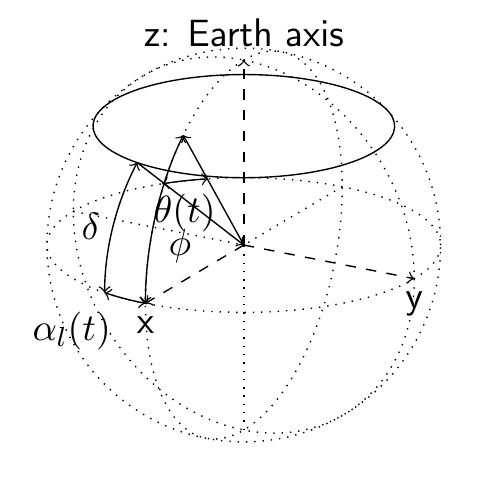}
\par\end{centering}

\caption{\label{fig:Scanning an area}\textit{Longitudinal dependencies for
Amsterdam Science Park follow the distributions in circles for several
declinations. ($\delta\protect\neq\phi$).}}
\end{figure}

\begin{figure}
\noindent \begin{centering}
\includegraphics[width=8cm]{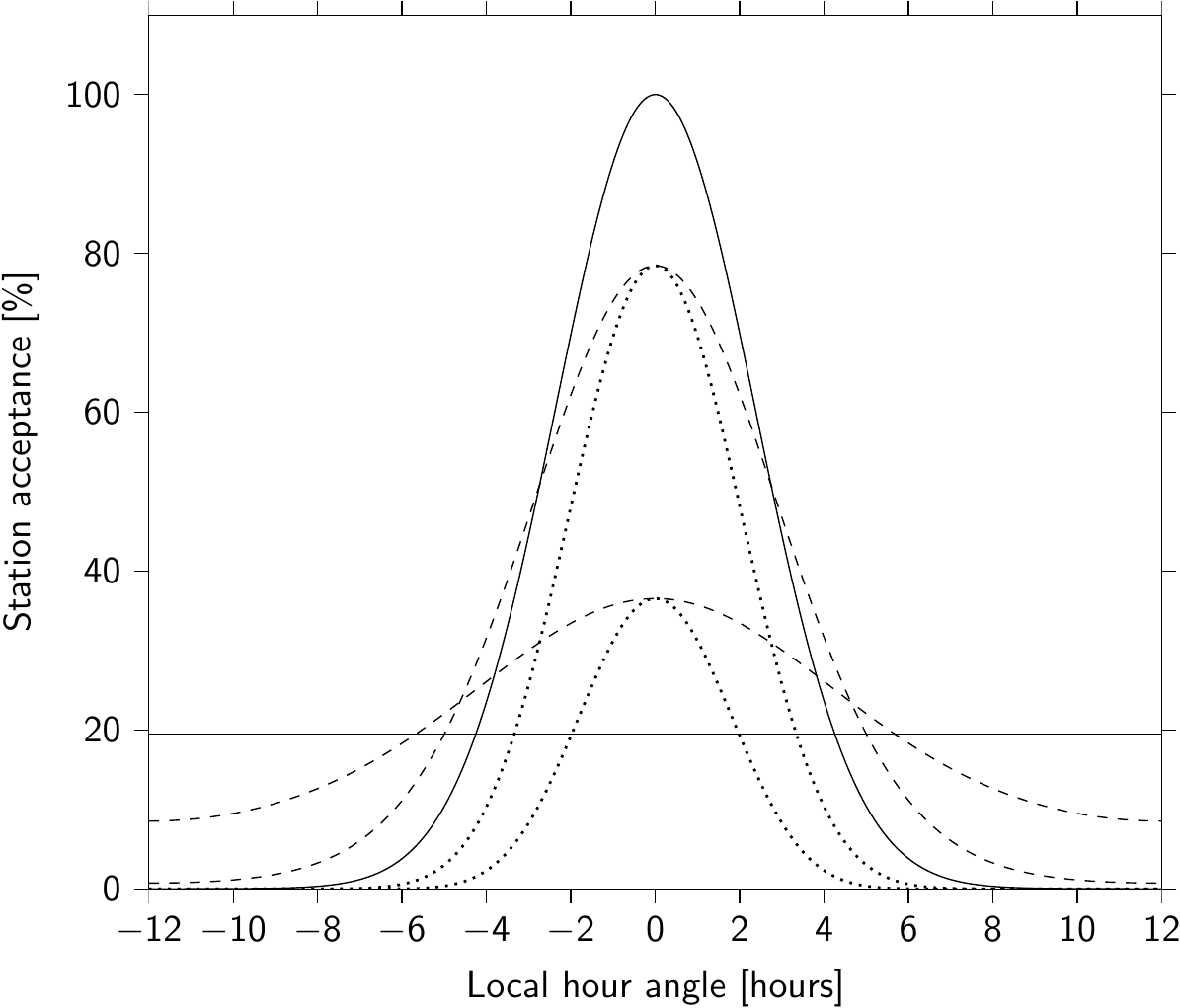}
\par\end{centering}

\caption{\textit{\label{fig:DeclinationDependentDistribution}Declination dependent
distributions for Science Park Amsterdam.}\noun{ }\textit{The distributions
for $\delta=\phi=52.35^{\mathrm{o}}$ (solid) through the zenith,
$\delta=\phi-15^{\mathrm{o}}$ and $\delta=\phi-30^{\mathrm{o}}$ to the south (dotted)
and $\delta=\phi+15^{\mathrm{o}}$ and $\delta=\phi+30^{\mathrm{o}}$ to the north
(dashed) are shown as function of the local hour angle. The horizontal
line corresponds with $\delta=90^{\mathrm{o}}$ (Polaris).}}
\end{figure}

EAS emerging from areas with higher or lower declinations then the latitude are detected too. 
The time windows of these distributions are analysed using the Amsterdam Science Park cluster 
at the center of the HiPARC experiment.
 
The Amsterdam Science Park cluster, located around $(\lambda,\phi)=(\ang{4.95}\mathrm{E}, \ang{52.35}\mathrm{N})$,
contains 11 stations, each using 4 scintillators, scattered over an area of $\SI{0.15}{\kilo\meter\squared}$. 
One of the stations is located in the lobby of the Nikhef building. All other stations are located on 
roof tops. 

As shown in fig. \ref{fig:DeclinationDependentDistribution} the distribution is flat for 
$\delta=90^{\mathrm{o}}$. Moving down in steps from $\delta=\phi+30^{\mathrm{o}}$ 
to $\delta=\phi-30^{\mathrm{o}}$ the width of the time window decreases considerably.

\section{The latitudinal dependency of a detected celestial area}

As shown in fig. \ref{fig:DeclinationDependentDistribution} the longitudinal
distribution of a station with a fixed latitude $\phi$ largely depends on the declination $\delta$. 

The location of a possible cosmic ray hot spot can be defined in the right ascension/declination 
$\left(\alpha,\delta\right)$-frame. The daily number of detected particles in EASs, initiated by these 
cosmic rays, is proportional to the integral of the acceptance (eq. \ref{eq:rossi-local}) over a sidereal 
day as long as $\cos\left(\theta\right)\geq0$, or $\delta+\phi>90^{\mathrm{o}}$. 

\begin{figure}[h]
\noindent \begin{centering}
\includegraphics[width=8cm]{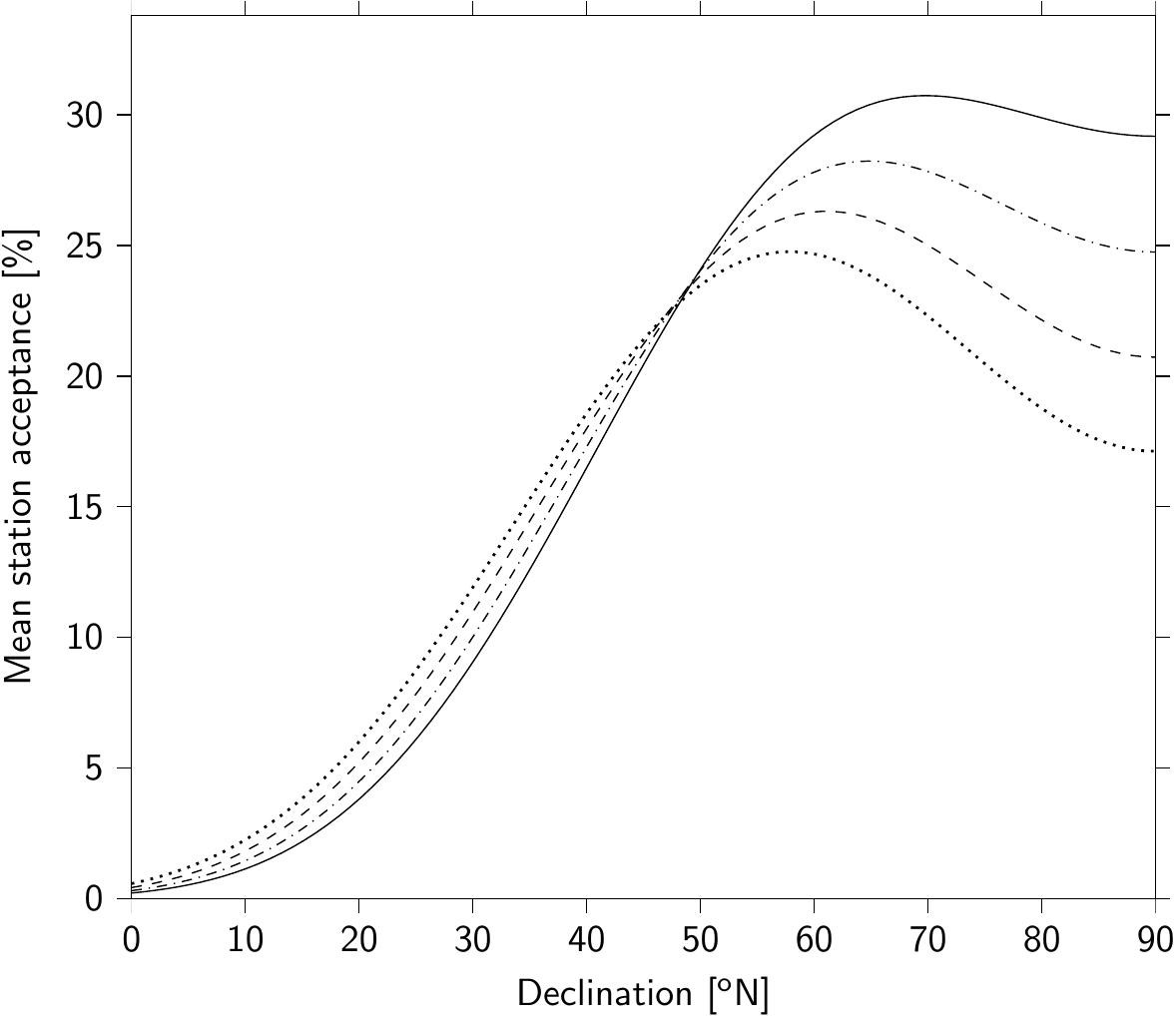}
\par\end{centering}

\caption{\label{fig:The-latitudinal-distribution}\textit{The mean acceptance for
an area with a declination is calculated by integrating eq. \ref{eq:rossi-local}
over a sidereal day. Shown are the acceptancies for $\phi=51^{\mathrm{o}}$
(dotted), $53^{\mathrm{o}}$ (dashed), $55^{\mathrm{o}}$ (dot dashed)
and $57^{\mathrm{o}}$ (solid). The equator has a declination $\delta=0^{\mathrm{o}}$,
Polaris has $\delta=90^{\mathrm{o}}$.}}
\end{figure}

The integral depends on the parameters $\delta$ and $\phi$. The
resulting distributions, found by integration of eq. \ref{eq:rossi-local} over $\alpha_{l}$, are shown as a 
function of the declination $\delta$ for different values of the lattitude $\phi$ in fig. 
\ref{fig:The-latitudinal-distribution}. During an entire day the HiSPARC experiment detects cosmic rays 
arriving from around Polaris ($\delta=\ang{90}$).

\section{Experimental verification}

\begin{figure}[h]
\noindent \begin{centering}
\includegraphics[width=12cm]{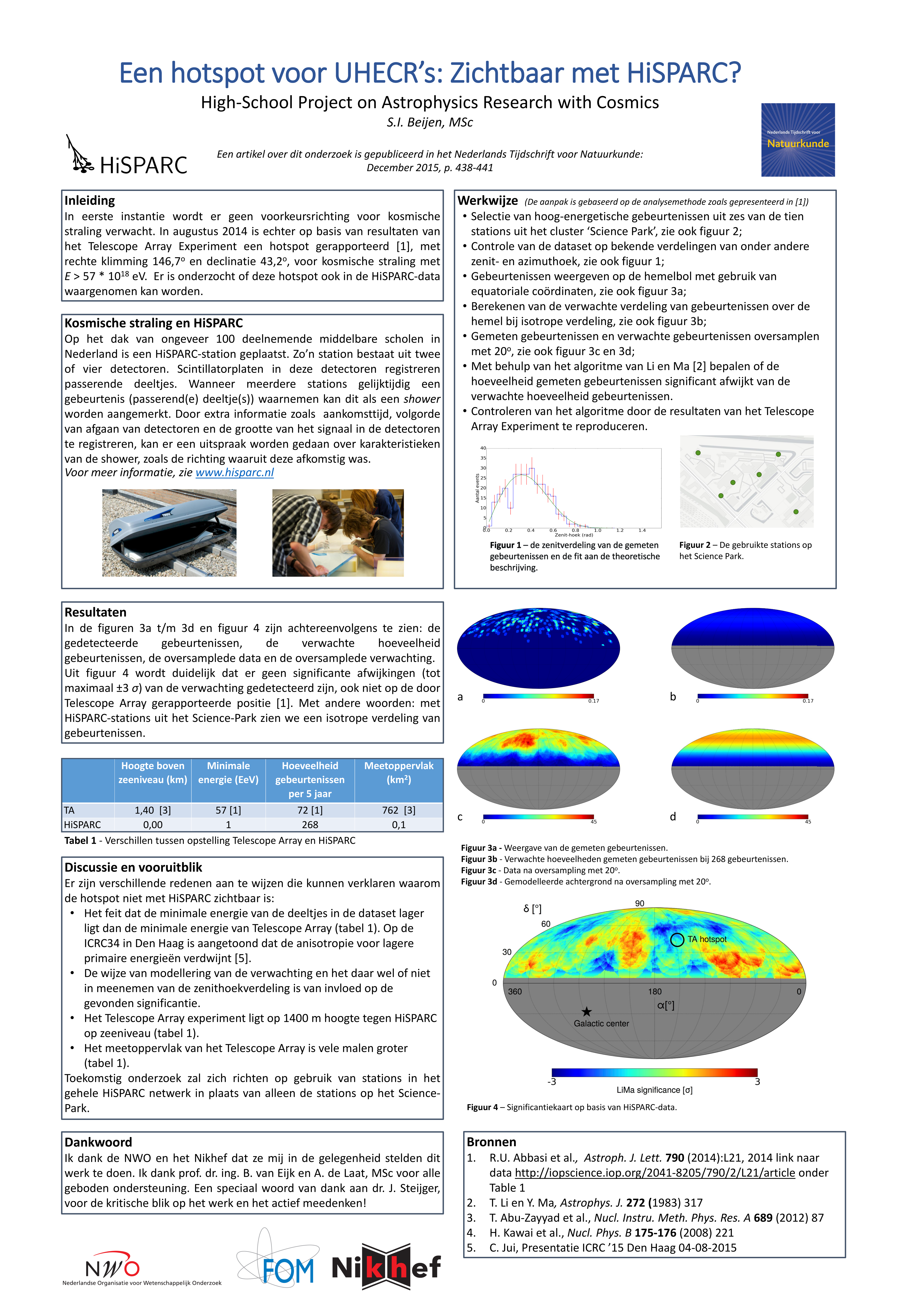}
\par\end{centering}

\caption{\label{fig:hotspot}\textit{A Mollweide view of the measured flux of cosmic rays with $E>1\si{EeV}$ 
by HiSPARC scaled by the acceptance, from \cite{SB-HS-2015}. A significance of $3\sigma$ 
can be interpreted as maximum difference of .3\%. The HiSPARC measurements and the 
acceptance of fig. \ref{fig:The-latitudinal-distribution} agree.}}

\end{figure}

The distribution of fig. \ref{fig:The-latitudinal-distribution} has been used in a HiSPARC survey of 
cosmic ray hotspots \cite{SB-HS-2015} with $E>1\si{EeV}$. In fig. \ref{fig:hotspot}\footnote{'TA hotspot' 
indicates the location of the hotspot as measured by Telescope Array \cite{Abbasi-2014} for $E>57\si{EeV}$.}  
the flux, as measured by the Amsterdam Science Park, is scaled using the acceptance. A significance of 
$3\sigma$ can be interpreted as maximum difference of .3\%. So the flux measured by HiSPARC and the 
acceptance of fig. \ref{fig:The-latitudinal-distribution} are in close agreement. 

\section{Conclusion}

\begin{figure}[h]
\noindent \begin{centering}
\includegraphics[width=10cm]{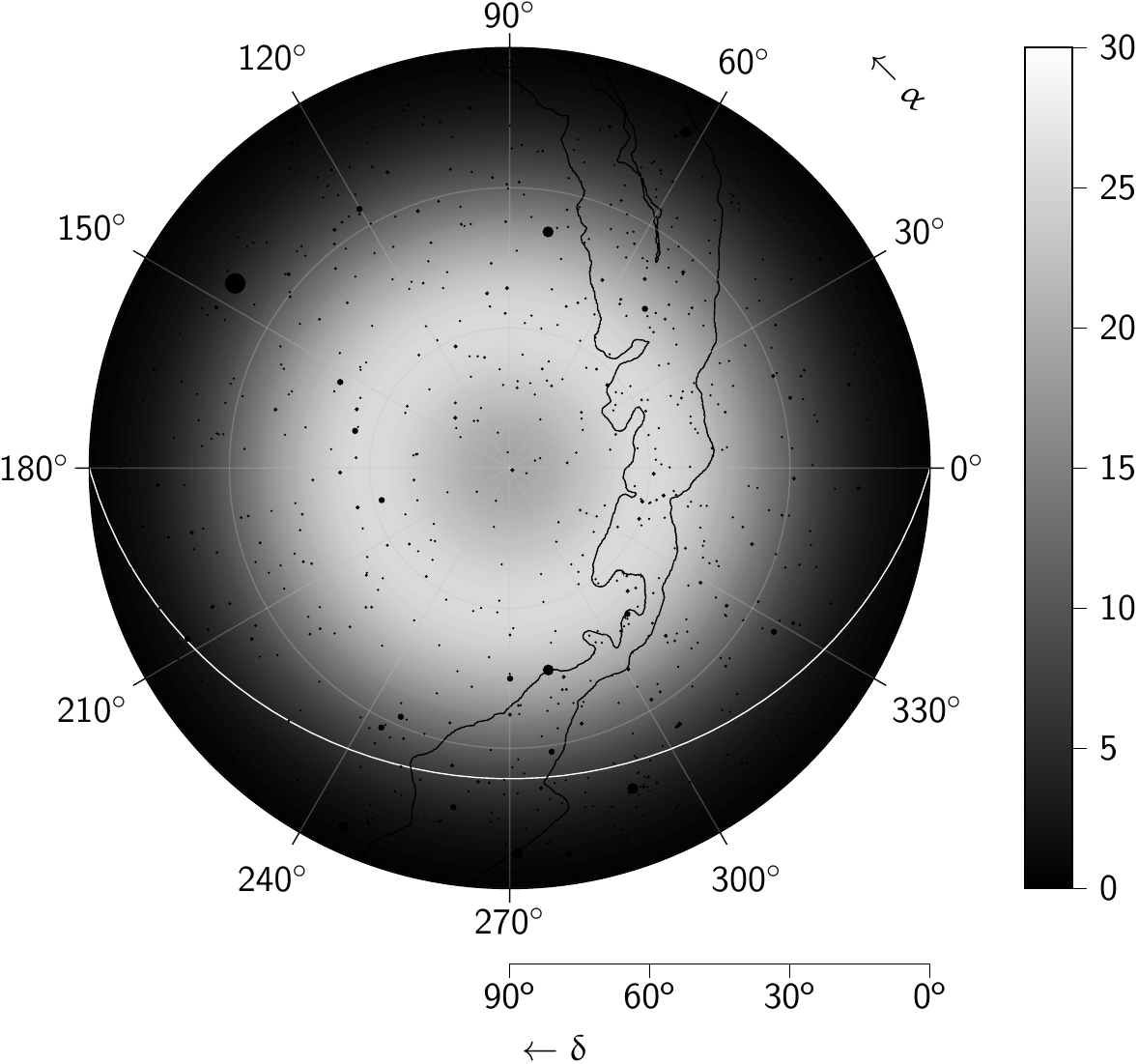}
\par\end{centering}

\caption{\label{fig:sky}\textit{The mean field of view of the HiSPARC Science Park cluster during a day. The Suns 
trajectory / ecliptic is shown as a white curve. Both stars and the edges of the Milky Way are shown in black. 
The acceptance is shown in grey-scales from 0\% (black) to 30\% (white).}}

\end{figure}

The HiSPARC experiment has a symmetric field of view around Polaris. In fig. \ref{fig:sky} the Suns 
trajectory / ecliptic is shown as a white curve. The acceptance is shown in grey-scales from 0\% (black) 
to 30\% (white) for the Science Park cluster. Both the borders of the Milky Way and the stars are shown 
in black. Ursa major is on the left side. 

The solar magnetic field influences the trajectory of the charged
cosmic ray primaries in a high degree within an angle of $45^{\mathrm{o}}$
measured from the direction of the Sun, a point on the ecliptica. 
The geometry of the HiSPARC experiment, with distances between detection 
stations of up to 1000km, facilitates the detection of the Gerasimova-Zatsepin effect. 

The magnetic field of the Sun effects the distance between the resulting simultaneous 
parallel EAS caused by a spallation \cite{MTW-1999}. During winter the Sun is always 
outside the view of the HiSPARC experiment. During a part of the day in summer the magnetic 
field of the Sun causes distances between simultaneous parallel EAS larger then the area 
covered by the HiSPARC experiment.

\newpage

\printbibliography

\end{document}